\documentclass[aps,prb,reprint,showpacs,superscriptaddress]{revtex4-2}
\usepackage{bm,amsfonts,amssymb,amsmath,mathrsfs}
\usepackage{graphicx,subfigure}
\usepackage{float}
\usepackage{xcolor}
\usepackage[colorlinks,linkcolor=blue,citecolor=blue,urlcolor=blue]{hyperref}

\usepackage{bbold}
\usepackage{color}

\usepackage{amsfonts,amsmath}
\usepackage{epsfig,amsmath}
\usepackage{graphicx}
\usepackage{dcolumn}
\usepackage{bm}
\usepackage{bbold}
\newcommand{\beq}{\begin{equation}}
\newcommand{\eeq}{\end{equation}}
\newcommand{\beqa}{\begin{eqnarray}}
\newcommand{\eeqa}{\end{eqnarray}}

\newcommand{\la}{\langle} 
\newcommand{\ra}{\rangle}

\makeatletter
\newcommand*{\revision}{\@ifnextchar\bgroup{\revision@}{\color{red}}}
\newcommand*{\revision@}[1]{{\textcolor{black}{#1}}}
\makeatother

\begin{document}
	
\title{Theory of Magnon Purcell Effect in Cavity Magnonic Systems}
\author{Guogan Zhao}
\affiliation{Department of Physics, and Center for Quantum Science and Engineering, Stevens Institute of Technology, Hoboken, NJ 07030, USA}
\author{Yong Wang}
\affiliation{School of Physics, Nankai University, Tianjin 300071, China}
\author{Xiao-Feng Qian}\email[]{xqian6@stevens.edu}
\affiliation{Department of Physics, and Center for Quantum Science and Engineering, Stevens Institute of Technology, Hoboken, NJ 07030, USA}

\begin{abstract}
We conduct a systematic analysis of cavity effects on the decay dynamics of an open magnonic system. \revision {The Purcell effect on the magnon oscillator decay is thoroughly examined for both driven and non-driven scenarios with realistic parameter and initial conditions. Analytical conditions are determined to distinguish between strong and weak coupling regimes, corresponding to oscillatory and pure decay behaviors respectively. Additionally, our theory also predicts the decay of the photon mode within the cavity-magnonic open system, demonstrating excellent agreement with existing experimental data. Our findings and methodologies may provide valuable insights for advancing research in cavity magnonic quantum control, quantum information processing, and the development of magnonic quantum devices.}
\end{abstract}
\maketitle






\section {Introduction} 
Cavity magnonic hybrid systems are considered a promising platform for information transduction in quantum science and engineering \cite{zarerameshti2022cavity}. They have garnered significant attention in recent years due to the fact that magnons can coherently exchange quantum information with various types of physical systems, mediated by confined electromagnetic fields within cavities or resonators. For instance, magnons have been shown to be able to couple directly with microwave cavity photon modes in both strong and ultra-strong coupling regimes \cite{bai2015spin, bourhill2016ultrahigh, goryachev2014highcooperativity, kostylev2016superstrong, tabuchi2014hybridizing, zhang2014strongly, wang2018bistability}, and with optical cavity photon modes via magneto-optical interactions \cite{haigh2016tripleresonant, osada2016cavity, osada2018brillouin, zhang2016optomagnonic, braggio2017optical, hisatomi2016bidirectional, sharma2018optical}. Magnons can also couple indirectly with superconducting circuit qubits (SC-qubits) \cite{tabuchi2015coherent, wolski2020dissipationbased, xu2023quantum, lachance-quirion2020entanglementbased} and mechanical vibration (phonons) modes \cite{potts2021dynamical, shen2022coherent, shen2023polaromechanics, zhang2016cavity, li2018magnonphotonphonon} via microwave cavities. Apparently, electromagnetic cavities play a crucial role in these processes, as they host confined electromagnetic field modes, which directly interact with the magnons in a controllable manner. However, in practical scenarios, both the cavities and the quantum systems within them are inevitably subject to environmental interactions, resulting decoherence \cite{zurek2003decoherence} or disentanglement \cite{yu2009sudden}. Quantum control via cavities is considered to be an efficient way of maintain these quantum resources \cite{lachance-quirion2019hybrid, landgraf2023fast, yuan2022quantum}. Therefore, it becomes particularly important to understand the cavity effects on magnon system decay. Experimental observation of photon mode decay enhancement effect has been observed in a cavity magnonic system \cite{zhang2014strongly}. However, such an effect, theory or experiment, for magnons is yet to be explored.





Purcell effect is one of the most prominent phenomena in open quantum systems involving cavities \cite{deliberato2014lightmatter, plankensteiner2019enhanced, rybin2016purcell, sete2014purcell}. It refers to the enhancement or suppression of an emitter's spontaneous emission rate by modifying its surrounding electromagnetic environment \cite{purcell1946spontaneous}. This effect has broad applications, such as enhancing single-photon source emission rate \cite{englund2005controlling, houck2007generating, munnix2009modeling, giesz2013influence}, improving laser efficiency \cite{breeze2015enhanced, romeira2018purcell}, increasing efficiency in solar cells \cite{lee2023gigantic}, etc. Previous research on the Purcell effect has primarily focused on two-level systems. \revision{More recently, there has been an extension to effective three-level systems \cite{nation2023multilevel,agarwal2024control}. However, systematic theoretical investigations of arbitrary multi-level systems remain largely unexplored. This gap is particularly significant for magnon systems, which exhibit oscillator multi-level decay dynamics, underscoring a critical area that has yet to be fully understood.}



In this work, we will address the two aforementioned gaps together by investigating a cavity magnonic open system. Specifically, we will (1) establish a systematic theory of cavity effects on open system magnon decay, and (2) extend the study of Purcell effect from few-level systems to the arbitrary multi-level magnon oscillator case. Decay dynamics of magnon excitations is analyzed in detail for both driven and non-driven cases, identifying specific weak coupling conditions under which Purcell effect is observed. Our theory also predicts the decay behavior of the cavity's average photon number in perfect agreement with existing experimental data, confirming our theoretical framework. Our findings offer valuable insights in the quantum control of cavity magnonic systems, thus beneficial to practical applications of various cavity magnon quantum tasks and devices.





\section{Purcell effect without drive} 

\begin{figure}[htbp]
    \centering
\includegraphics[width=\linewidth]{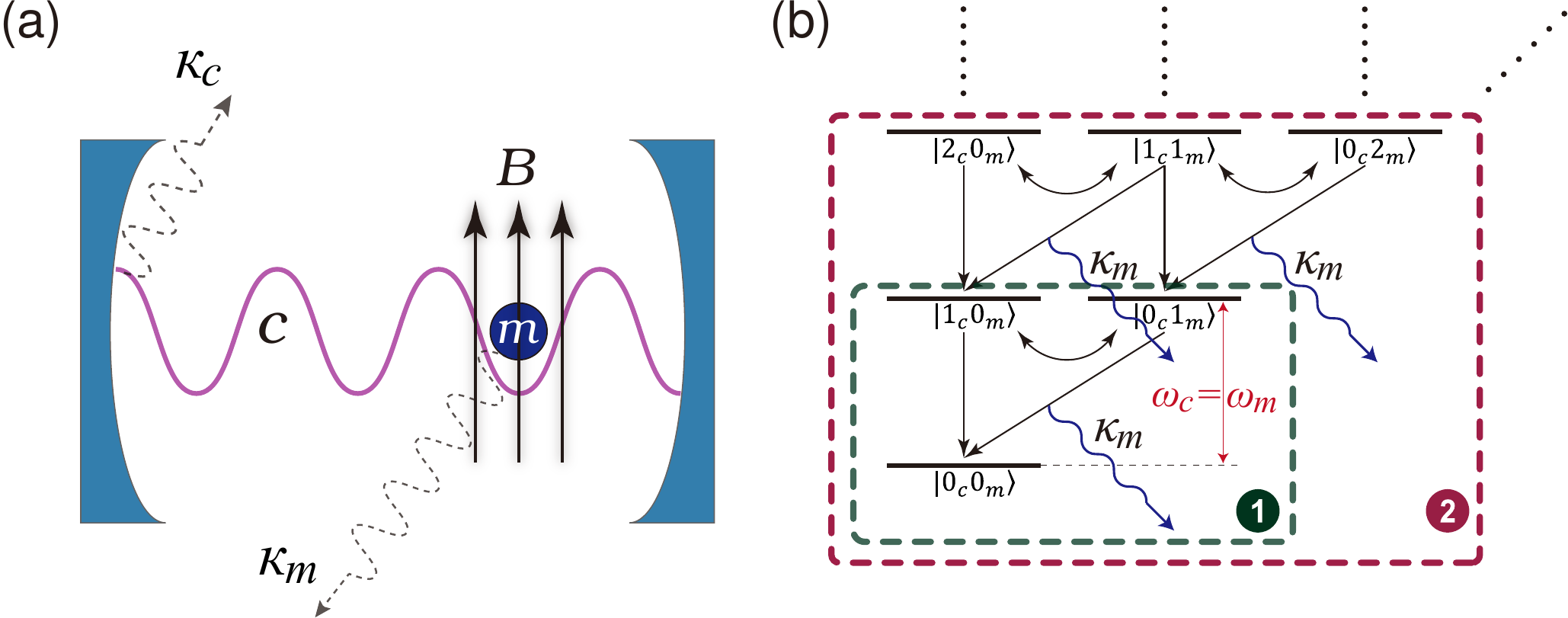}
    \caption{(a) Schematic diagram of cavity magnon system with a bias magnetic filed $B$ applied to the magnet sphere. The microwave cavity mode $c$ and magnon mode $m$ decays at a rate of $\kappa_c$ and $\kappa_m$ respectively. (b) Schematic diagram of magnon decay transitions from all energy levels. Enclosed areas 1 and 2 illustrate examples of all contributed decay transitions involving one and two excitations, respectively.}\label{schematic}
\end{figure}

We start by considering the commonly studied cavity magnonic scenario, see illustration in Fig.~\ref{schematic}(a), where a magnet (yttrium iron garnet sphere), placed inside a microwave cavity, is coherently coupled to the cavity mode via magnetic dipole interaction. It is assumed that only the Kittel mode \cite{kittel1948theory} of the magnet is excited by the magnetic component of the microwave field and a bias magnetic field $B$ is applied on the magnet. The Hamiltonian, under rotating-wave approximation (RWA), of the system reads \cite{zhang2014strongly,zarerameshti2022cavity}
\begin{equation}
H=\hbar\omega_{c}c^{\dag}c+\hbar\omega_{m}m^{\dag}m+\hbar g(c^{\dag}m+ m^{\dag}c),\label{Ham1}
\end{equation}
where $c^{\dag}~(c)$ and $m^{\dag}~(m)$ are the creation (annihilation) operators of cavity photons with frequency $\omega_c$ and magnons with frequency $\omega_m$ respectively, $g$ is the coupling constant between photon and magnon modes. The magnon frequency is mediated by a bias magnetic field $B$, i.e., $\omega_m=\gamma B$, where $\gamma$ is the gyromagnetic ratio.

To analyze the environment-induced decay process, we follow the standard open quantum system approach with the Markovian approximation \cite{breuer2002theory, ficek_quantum_2014}. Then the time evolution of the reduced system state $\rho$ is described by the master equation
\begin{equation}
\begin{aligned}
\frac{d\rho}{dt}={}&\frac{1}{i\hbar}[H,\rho]-\kappa_{c}(c^{\dag}c\rho-2c\rho c^{\dag}+\rho c^{\dag}c) \\
&-\kappa_{m}(m^{\dag}m\rho-2m\rho m^{\dag}+\rho m^{\dag}m),\label{masterequ}
\end{aligned}
\end{equation}
where $\kappa_c$ and $\kappa_m$ are the intrinsic decay rates of the cavity photons and magnon modes respectively. \revision{Note that master equation \eqref{masterequ} is derived within RWA, which requires $g\ll \omega_c, \omega_m$ and the validity of this equation requires the thermal occupations of the environment are negligible compared to coherent excitations in the system \cite{zhao2021driven}.}

Unlike traditional two-level system, the multilevel nature of the magnon oscillator decay is composed of a structured multiple level transitions. As an illustration, Fig.~\ref{schematic} (b) shows the structure of all relevant transitions for single and two excitations with the enclosed areas 1 and 2 respectively. For example, in the one excitation case, only the one-photon (and zero-magnon) state $|1_c 0_m\rangle$, and the one-magnon (and zero-photon) state $| 0_c 1_m\rangle$ are involved \cite{luo2021nonlocal,luo2024optically}. Similarly, for the two-excitation case, one will need to involve three more states, i.e., $|2_c0_m\ra$, $|1_c1_m\ra$, $|0_c2_m\ra$. However, as the number of excitations increases, it becomes more and more complex to include all level transitions. Moreover, in most cases, one has no prior knowledge of the number of excitations in the systems. Therefore, such a structured treatment is in general impractical. 

To overcome this issue, we approach it with a direct analysis of the equations of motion of all average excitation numbers, i.e.,  $\operatorname{Tr}[\hat{O}\rho]$ where $\hat{O}=c^{\dag} c, m^{\dag} m, c^{\dag} m$ and $m^{\dag} c$ (see details in Appendix \ref{AppA}). Such a treatment is a coherent inclusion of all level transitions in a single dynamical process. With a fixed arbitrary initial magnon excitation number, the dynamical equations for magnon (and photon) numbers can be achieved via the master equation \eqref{masterequ}, and is given as 
\begin{equation}
\frac{d}{d t}
\left(\begin{array}{c}
\left\langle c^{\dag} c\right\rangle \\
\left\langle m^{\dag} m\right\rangle \\
\left\langle c^{\dag} m\right\rangle \\
\left\langle m^{\dag} c \right\rangle
\end{array}\right)=
\mathcal{C}
\left(\begin{array}{c}
\left\langle c^{\dag} c\right\rangle \\
\left\langle m^{\dag} m\right\rangle \\
\left\langle c^{\dag} m\right\rangle \\
\left\langle m^{\dag} c \right\rangle
\end{array}\right).\label{NumEq}
\end{equation}
Here the coefficient matrix $\mathcal{C}$ is given as 
\begin{equation}
\begin{aligned}
\mathcal{C}=\left(\begin{array}{cccc}
-2 \kappa_c & 0 & -i g & i g \\
0 & -2 \kappa_m & i g & -i g \\
-i g & i g & -i \Delta-\kappa & 0 \\
i g & -i g & 0 & i \Delta-\kappa
\end{array}\right)
\end{aligned},\label{coefficient1}
\end{equation}
with detuning $\Delta=\omega_m-\omega_c$ and $\kappa=\kappa_c+\kappa_m$. There are four eigenvalues of the coefficient matrix, i.e.,
\begin{equation}
\begin{aligned}
\Gamma_1={}& -\kappa -\sqrt{(\delta-\delta')/2},  \quad \Gamma_2= -\kappa +\sqrt{(\delta-\delta')/2}, \\
\Gamma_3={}& -\kappa -\sqrt{(\delta+\delta')/2}, \quad \Gamma_4= -\kappa +\sqrt{(\delta+\delta')/2}, 
\end{aligned}\label{FundamentalChannels}
\end{equation}
where $\delta=d^2-4g^2-\Delta^2$, $\delta'=\sqrt{ \delta^2+4\Delta^2d^2}$, and $d=\kappa_c - \kappa_m$. These four eigenvalues correspond to four different decay channels of the system. That is, one can diagonalize the coefficient matrix $\mathcal{C}$ with matrix $U$, such that $U\mathcal{C}U^{-1}$ is diagonal. Then an eigenmode equation can be achieved from (\ref{NumEq}), i.e., $\la I_i(t)\ra=\la I_i(0)\ra e^{\Gamma_i t}$, $i=1,2,3,4$, where $\la I_i(t)\ra$ are the eigen modes transformed from of the average excitations $\la c^{\dag} c\ra$, $\la m^{\dag} m\ra$, $\la c^{\dag} m\ra$, $\la m^{\dag} c\ra$ through the $U$ matrix, and are known as the cavity magnon polaritons \cite{match2019transient, zhao2021driven}.

\begin{figure}[h]
    \centering
\includegraphics[width=\columnwidth]{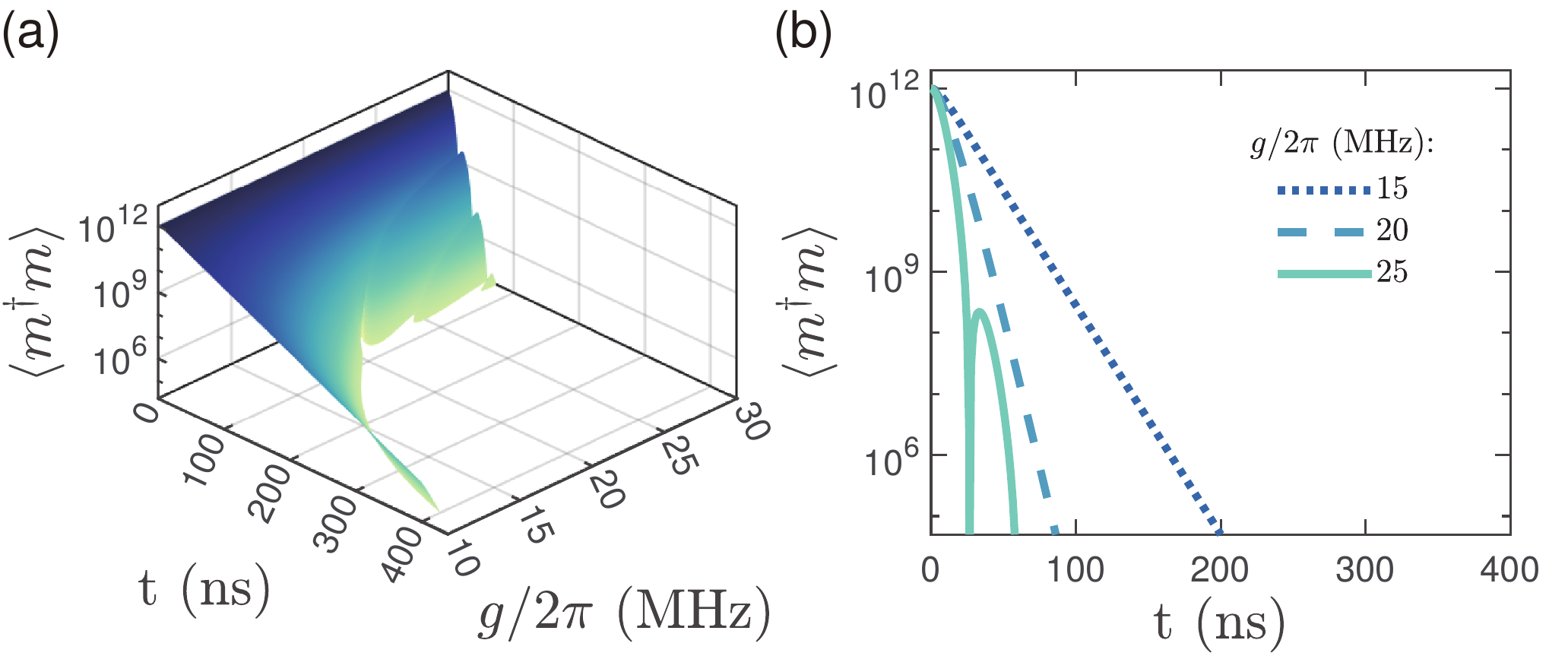}
    \caption{(a) Logarithm illustration of magnon number decay for continuously varying coupling rates. (b) Magnon decay for three typical coupling values of $g$. Experimentally feasible parameters are adopted to be $\omega_c/2\pi=\omega_m/2\pi=5$ GHz, $\kappa_c/2\pi=40.1$ MHz and $\kappa_m/2\pi=0.1$ MHz.}\label{decay_with_g}
\end{figure}

When the magnon mode is resonant with the cavity mode, $\Delta=\omega_m-\omega_c=0$, two of the four decay channels degenerate, i.e., $\Gamma_1=\Gamma_2=-\kappa$ when $\delta \ge 0$ or $\Gamma_3=\Gamma_4=-\kappa$ when $\delta < 0$. Therefore, the solutions of Eq.~\eqref{NumEq} can be obtained with three decay channels, i.e.,
\begin{equation}
\begin{aligned}
\left\langle m^{\dagger} m\right\rangle (t) =&{}N_m\left[ a_0 e^{-\kappa t} + a_1 e^{-\left(\kappa+\sqrt{\delta}\right)t}+ a_2 e^{-\left(\kappa-\sqrt{\delta}\right)t}\right], \\
\left\langle c^{\dagger} c\right\rangle(t) =&{}\frac{1}{2}N_m a_0 \left[2 e^{-\kappa t}-e^{-\left(\kappa+\sqrt{\delta}\right)t}-e^{-\left(\kappa-\sqrt{\delta}\right)t}\right],
\end{aligned}\label{solutions1}
\end{equation}
where $a_0=-2 g^2/\delta$, $a_1=(d-\sqrt{\delta})^2/4\delta$ and $a_2=(d+\sqrt{\delta})^2/4\delta$. Here we have assumed initially at time $t=0$, there are $N_m$ magnons and zero photons, i.e., $\langle m^{\dag} m \rangle=N_m$ and $\langle c^{\dag} c \rangle=\langle c^{\dag} m \rangle=\langle m^{\dag} c \rangle=0$.

Fig.~\ref{decay_with_g} (a) illustrates the time evolution of the average magnon number $\langle m^{\dagger} m\rangle$ over a broad range of coupling strengths with experimentally feasible parameters \cite{zhang2014strongly}: $\omega_c/2\pi=\omega_m/2\pi=5$ GHz, $\kappa_c/2\pi=40.1$ MHz and $\kappa_m/2\pi=0.1$ MHz. Obviously, it displays a transition from an pure decay to an oscillatory decay as the coupling rate increases. Such a transition can be explicitly determined by coupling strength $g$ through the parameter $\delta=d^2-4g^2-\Delta^2$. When $\delta \ge 0$, i.e.,  $0<g\le(\kappa_c-\kappa_m)/2$, all three decay channels in Eq.~\eqref{solutions1} contain real parameters, thus demonstrating pure decays. When $\delta < 0$, i.e., $g>(\kappa_c-\kappa_m)/2$, two of the three channels in \eqref{solutions1} contain imaginary terms, thus showing oscillatory behaviors. 

This transition point naturally defines the weak coupling regime, $0<g\le(\kappa_c-\kappa_m)/2$, with a pure decay process. This is the region of interest here for the investigation of Purcell effect. This definition coincides with previous studies in Ref.~\cite{harder2017topological}.

\begin{figure}[t]
    \centering    \includegraphics[width=\columnwidth]{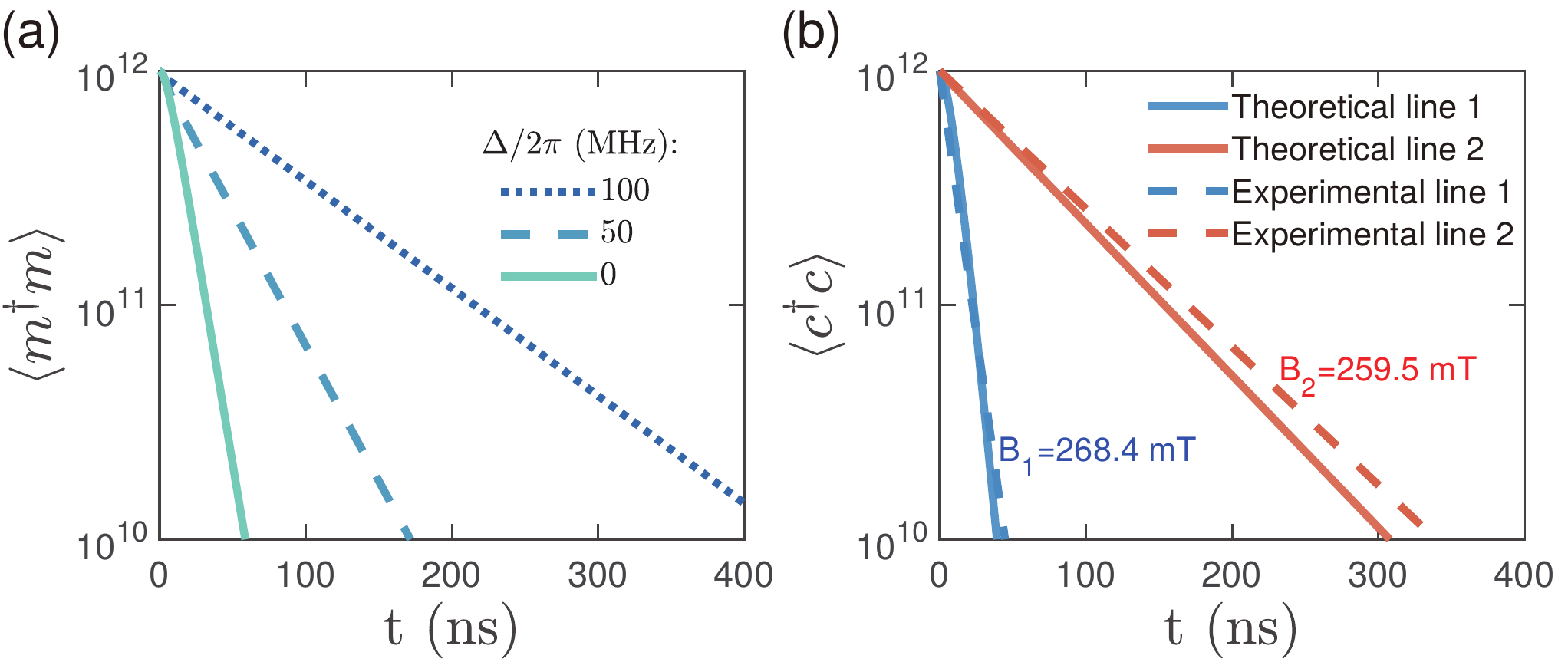}
    \caption{(a) Decay of magnons for detunings $\Delta=$0, 50 and 100 MHz with $g/2\pi=15$ MHz. The remaining parameters are the same as those in Fig.~\ref{decay_with_g}. (b) Decay of cavity photons with experimental parameters: $\omega_c/2\pi=7.58$ GHz, $g/2\pi=15$ MHz, $\kappa_c/2\pi=1.08$ MHz and $\kappa_m/2\pi=32$ MHz. The dashed lines are fitted data from the experimental work \cite{zhang2014strongly} and the solid lines are our theoretical results from \eqref{solutions1}.}\label{decay_with_detuning}
\end{figure}


Examples of specific magnon decay at three typical coupling strengths is provided in Fig.~\ref{decay_with_g} (b). Oscillatory behavior is shown by the solid curve when $\delta <0$. In the weak coupling regime $\delta\ge 0$, the dashed and dotted curves show slow decay in the initial short-time window due to the negative coefficient $a_0$ in Eq.~\eqref{solutions1}. \revision{As time progresses, magnons decay faster through the first two channels of Eq.~\eqref{solutions1} and the third channel due to the fact that $\kappa -\sqrt{\delta} \le \kappa \le \kappa +\sqrt{\delta}$. Thus, in the long-time regime ($t\rightarrow \infty$), the slowest decay channel $e^{-(\kappa -\sqrt{\delta})t}$ will dominate the dynamics and become the main decay process for the entire system.} More importantly, the dotted and dashed curves show directly Purcell effect, i.e., the increase of coupling strength will significantly increase the decay rate with a Purcell Factor of $(\kappa-\sqrt{\delta})/2\kappa_m$ compared with the intrinsic decay rate at zero coupling $g=0$. In this example, the dotted and dashed curves represent a Purcell Factor of 69 and 201, respectively.



When adjusting the bias magnetic field $B$, the magnon mode can be detuned from the cavity mode, i.e., $\Delta>0$. Then the dynamical behaviors of magnon and cavity photon numbers can be achieved numerically through \eqref{NumEq}. Fig.~\ref{decay_with_detuning} (a) illustrates the decay of magnon numbers for three different detunings. Apparently, detuning will suppress the Purcell effect as the increase of the $\Delta$ will decrease the decay rate. 

To compare our theoretical results with  experimental data, we compute from \eqref{NumEq} the decay of cavity photon number $\la c^{\dag} c\ra$ as was observed in Ref.~\cite{zhang2014strongly}. The dashed red and blue lines in Fig.~\ref{decay_with_detuning} (b) represent the experimental results from \cite{zhang2014strongly} for biased magnetic fields $B_1=268.4$ mT and $B_2=259.5$ mT, respectively. Our theoretical results for the same initial condition, $\la c^{\dag} c\ra=N_c=10^{12}$, $\la m^{\dag} m\ra=\la c^{\dag} m\ra=\la m^{\dag} c\ra=0$, are demonstrated by the corresponding solid red and blue lines. \revision{One notes that, our theoretical framework aligns perfectly with experimental data in the classical regime, validating its robustness. Importantly, this analysis extends to purely quantum states, such as Fock states or mode-entangled states, offering a pathway to study open system quantum magnon-photon dynamics, a relatively less explored domain in cavity magnonics.}



\section{Purcell effect with drive}
\label{section3}

Now we extend the analysis to the case when the system is driven by a coherent microwave field. Then the Hamiltonian is given as

\begin{equation}
\begin{aligned}
H={}&\hbar\omega_{c}c^{\dag}c+\hbar\omega_{m}m^{\dag}m+\hbar g(c^{\dag}m+ m^{\dag}c) \\
&+i\Omega(m^{\dag} e^{-i\omega_d t} - m e^{i\omega_d t}),\label{Ham2}
\end{aligned}
\end{equation}
where $\Omega$ and $\omega_d$ are the amplitude and frequency of the driving field, respectively. Then the dynamical equations for magnon and photon numbers can be achieved through master equation analysis, and are obtained as


\begin{equation}
\frac{d}{d t}
\left(\begin{array}{c}
\left\langle c^{\dag} c\right\rangle \\
\left\langle m^{\dag} m\right\rangle \\
\left\langle c^{\dag} m\right\rangle \\
\left\langle m^{\dag} c\right\rangle \\
\left\langle c^{\dag} \right\rangle ' \\
\left\langle c \right\rangle ' \\
\left\langle m^{\dag} \right\rangle ' \\
\left\langle m \right\rangle '
\end{array}\right)=
\mathcal{D}
\left(\begin{array}{c}
\left\langle c^{\dag} c\right\rangle \\
\left\langle m^{\dag} m\right\rangle \\
\left\langle c^{\dag} m\right\rangle \\
\left\langle m^{\dag} c \right\rangle \\
\left\langle c^{\dag} \right\rangle ' \\
\left\langle c \right\rangle ' \\
\left\langle m^{\dag} \right\rangle ' \\
\left\langle m \right\rangle '
\end{array}\right)
+
\left(\begin{array}{c}
0 \\ 0 \\ 0 \\ 0 \\ 0 \\ 0 \\ \Omega \\ \Omega
\end{array}\right),\label{NumEqD}
\end{equation}
where $\langle c^{\dag} \rangle '= \langle c^{\dag} \rangle e^{-i\omega_d t}$, $\langle c \rangle '= \langle c \rangle e^{i\omega_d t}$, $\langle m^{\dag} \rangle '= \langle m^{\dag} \rangle e^{-i\omega_d t}$, $\langle m \ra '= \langle m \rangle e^{i\omega_d t}$. Here $\mathcal{D}$ is an 8 dimensional coefficient matrix. Its eigen values define the eigen-mode decay channels. In addition to the four channels given in \eqref{FundamentalChannels}, the driving case contains four more channels
\begin{equation}
\begin{aligned}
& \Gamma_5=\frac{1}{2}\left(-\kappa +i\Delta' -\sqrt{\delta+2i\Delta d}\right), \\
& \Gamma_6=\frac{1}{2}\left(-\kappa +i\Delta' +\sqrt{\delta+2i\Delta d}\right), \\
& \Gamma_7=\frac{1}{2}\left(-\kappa -i\Delta' -\sqrt{\delta-2i\Delta d}\right), \\
& \Gamma_8=\frac{1}{2}\left(-\kappa -i\Delta' +\sqrt{\delta-2i\Delta d}\right), \label{8decaychannels}
\end{aligned}
\end{equation}
where $\Delta' = \omega_m +\omega_c -2 \omega_d$.


In general, the existence of the driving field $\omega_d$ introduces imaginary components in the first term of the eigenvalues $\Gamma_{5,6,7,8}$ above, which will lead to oscillatory behavior in the four additional channels. To analyze Purcell effect, we consider the representative resonant case with $\omega_c=\omega_m=\omega_d$, corresponding to a pure decay process. The non-resonant situation will lead to qualitatively similar effects. 

\bigskip

Here the coefficient matrix $\mathcal{D}$ is specifically given as 
\begin{widetext}
\begin{equation}
\begin{aligned}
\mathcal{D}=\left(\begin{array}{cccccccc}
-2 \kappa_c & 0 & -i g & i g & 0 & 0 & 0 & 0 \\
0 & -2 \kappa_m & i g & -i g & 0 & 0 & \Omega & \Omega \\
-i g & i g & -i \Delta-\kappa & 0 & \Omega & 0 & 0 & 0 \\
i g & -i g & 0 & i \Delta-\kappa & 0 & \Omega & 0 & 0 \\
0 & 0 & 0 & 0 & i (\omega_d-\omega_c)-\kappa_c & 0 & i g & 0 \\
0 & 0 & 0 & 0 & 0 & -i (\omega_d-\omega_c)-\kappa_c & 0 & -i g \\
0 & 0 & 0 & 0 & i g & 0 & i (\omega_d-\omega_m)-\kappa_m & 0 \\
0 & 0 & 0 & 0 & 0 & -i g & 0 & -i (\omega_d-\omega_m)-\kappa_m
\end{array}\right). \label{coefficient2}
\end{aligned}
\end{equation}
\end{widetext}


Due to a coherent drive, we assume initially the magnon mode is an coherent state with an average of $N_m$ magnons and there are zero cavity photons, i.e., $\langle m^{\dag} m \rangle=N_m$, $\langle m^{\dag} \rangle'=\langle m \rangle'=\sqrt{N_m}$ and $\langle c^{\dag} c \rangle=\langle c^{\dag} m \rangle=\langle m^{\dag} c \rangle=\langle c^{\dag} \rangle'=\langle c \rangle'=0$. We remark that other initial states, such as Fock states, will result in qualitatively similar decay behaviors determined by the decay channels \eqref{8decaychannels}. Then the magnon number dynamics can be obtained analytically with five dynamic channels
\begin{equation}
\begin{aligned}
\left\langle m^{\dagger} m\right\rangle (t) ={}& b_0 e^{-\kappa t} + b_1 e^{-\left(\kappa+\sqrt{\delta}\right)t}+ b_2 e^{-\left(\kappa-\sqrt{\delta}\right)t} \\
& + b_3 e^{-\frac{1}{2}\left(\kappa+\sqrt{\delta}\right)t} + b_4 e^{-\frac{1}{2}\left(\kappa-\sqrt{\delta}\right)t} \\
&+ \frac{\Omega^2 \kappa_c^2}{\left(g^2+\kappa_c \kappa_m\right)^2}.
\end{aligned}\label{solutions2}
\end{equation}
The reduction from eight to five channels is due to degeneracy caused by the resonant condition. Here the coefficients $b_0$, $b_1$, $b_2$, $b_3$ and $b_4$ are complex analytically expressions all containing the driving parameter $\Omega$, \revision{and are given in Appendix \ref{AppB}.}

\begin{figure}[htbp]
    \centering    \includegraphics[width=\columnwidth]{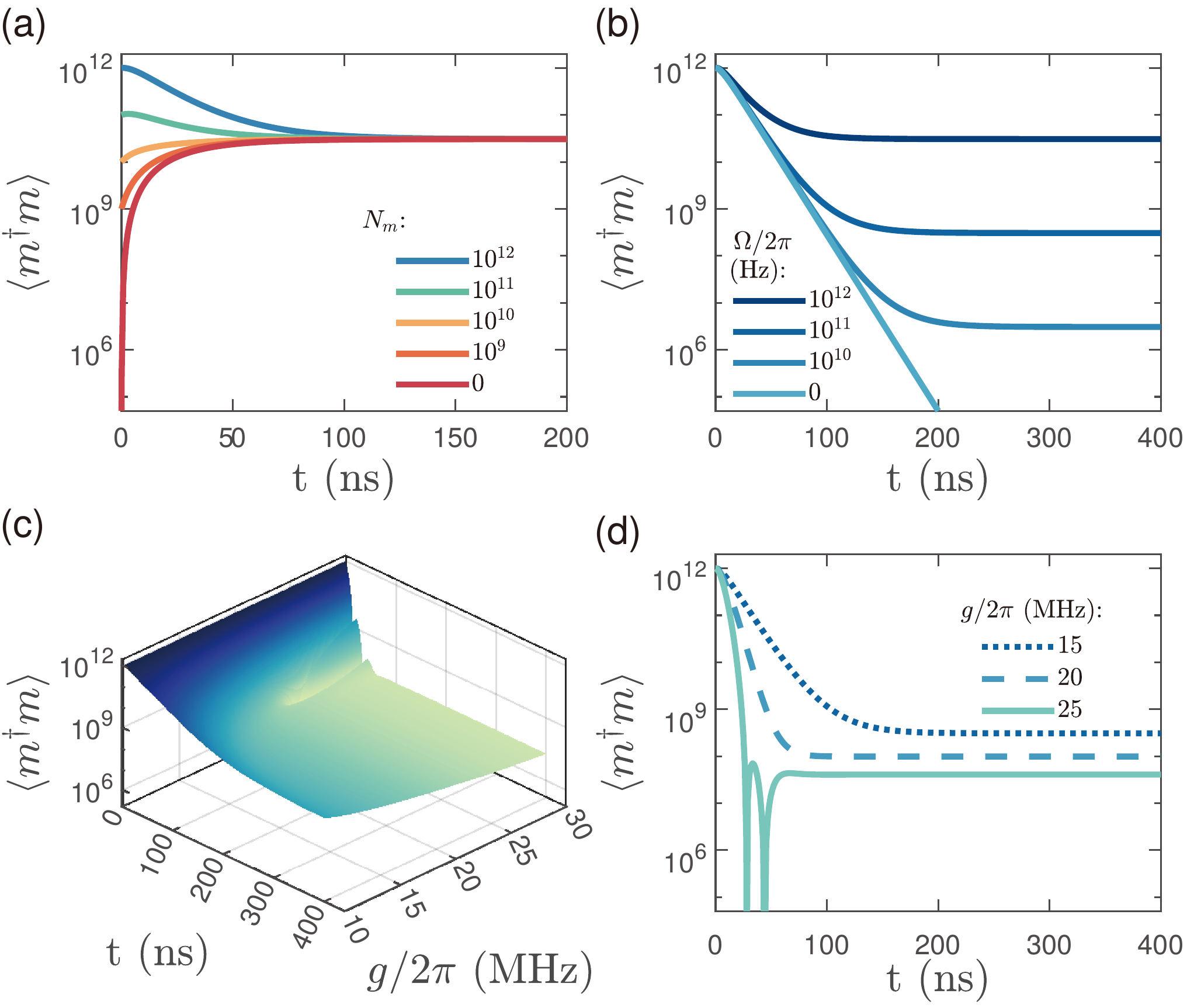}
    \caption{(a) Magnon decay with different initial average excitation $N_m$ and fixed drive strength $\Omega/2\pi=10^{12}$ Hz. (b) Magnon decay with fixed initial excitation number $N_m=10^{12}$ and different drive strength $\Omega$. (c) Magnons decay with for a continuously varying coupling $g$, fixed initial excitation $N_m=10^{12}$, and fixed drive strength $\Omega/2\pi=10^{11}$ Hz. (d) Typical examples of (c) for three values of $g$. We take $\omega_d/2\pi=\omega_c/2\pi=\omega_m/2\pi=5$ GHz and the other parameters are the same as in Fig.~\ref{decay_with_detuning} (a).}\label{decay_with_drive}
\end{figure}

Notably, as time $t \rightarrow \infty$, the existence of the drive will cause the dynamical evolution to approach to a steady state with a constant magnon number $N_{steady}=\Omega^2 \kappa_c^2/\left(g^2+\kappa_c \kappa_m\right)^2$, the last term of Eq.~\eqref{solutions2}. Obviously, when $\Omega=0$, it reduces to the no drive case \eqref{solutions1}, where the long-time magnon number simply decays to zero. Fig.~\ref{decay_with_drive} (a) illustrates exactly such a behavior for various initial magnon numbers $N_m$ at a fixed drive strength $\Omega/2\pi= 10^{12}$ Hz. Naturally, when $N_m >  N_{steady}$ it experiences a decay, and $N_m< N_{steady}$ leads to a magnon number increasing process. One notes that at the vicinity of the region when $N_m$ is slightly larger or equal than $N_{steady}$, the magnon number experiences an increase followed by a decay process. This is due to the fact that the signs of the coefficients $b_i$, $i=0,1,2,3,4$ depends on the initial magnon number $N_m$, which creates a competition between the five decay channels. 

Fig.~\ref{decay_with_drive} (b) illustrates the effect of driving strength $\Omega$ on magnon decay dynamics for a given initial excitation $N_m > N_{steady}$ and a coupling constant $g/2\pi=15$ MHz. One notes that the decrease of driving strength $\Omega$ will increase the decay rate, and it reduces exactly to the non-drive case when $\Omega=0$.


Fig.~\ref{decay_with_drive} (c) and (d) demonstrate how the magnon decay depends on the coupling strength $g$ for a fixed $\Omega/2\pi=10^{11}$ Hz and initial magnon number $N_m =10^{12}> N_{steady}\approx 10^8$. Apparently, the increase of coupling $g$ will significantly enhance the decay rate, manifestation of the Purcell effect, consistent with the non-drive case. Surprisingly, due to the last term in Eq.~\eqref{solutions2}, it is noted that a larger coupling rate will lead to a steady state with less magnons. The weak coupling region in this driving case is naturally defined through the decay channels of \eqref{solutions2}, i.e., $\delta\ge 0$ or $0<g\le(\kappa_c-\kappa_m)/2$, exactly the same as the non-drive case. For strong coupling $g>(\kappa_c-\kappa_m)/2$, oscillatory behavior arises as shown in Fig. ~\ref{decay_with_drive} (c). It is worth to note that the Purcell effect can also be extended to strong coupling cases in the short-time regime which permits a pure decay and the decay rate also increases as $g$ increases.    

We note that in non-resonant cases, the evolution of magnon number consistently exhibits oscillatory behavior. However, in the short-time regime, the effect of coupling strength on decay rate remains qualitatively similar, demonstrating the presence of the Purcell effect.

{
\section{Purcell effect of a steady state}
\label{section4}
While a magnon-only initial state is easy to analyze theoretically, it is difficult to prepare in practical experimental contexts. Here we extend the Purcell effect analysis to the realistic case of an initial steady state, of which both magnons and photons maintain a coherent state characteristic \cite{zhao2021driven}, i.e., $\left\langle m^{\dagger} m\right\rangle =\langle m^{\dag} \rangle'\langle m \rangle'$, $\left\langle c^{\dagger} c\right\rangle =\langle c^{\dag} \rangle'\langle c \rangle'$, $\left\langle m^{\dagger} c\right\rangle =\langle m^{\dag} \rangle'\langle c \rangle'$, $\left\langle c^{\dagger} m\right\rangle =\langle c^{\dag} \rangle'\langle m \rangle'$ (see also analysis in Appendix \ref{AppB}). In experimental implementation, such a steady state can be achieved by coherently driving the zero excitation cavity-magnon system. When a steady state
is prepared, one then has the following properties, $\langle m^{\dag} m \rangle=N_m$, $\langle c^{\dag} c \rangle=N_c$, $\langle m^{\dag} \rangle'=\langle m \rangle'=\sqrt{N_m}$, $\langle c^{\dag} \rangle'^*=\langle c \rangle'=i\sqrt{N_c}$ and $\langle c^{\dag} m \rangle^*=\langle m^{\dag} c \rangle=i\sqrt{N_m N_c}$. Then one can apply a new but weaker driving field to explore magnon number dynamics, which can be obtained as
\begin{equation}
\begin{aligned}
\left\langle m^{\dagger} m\right\rangle (t) ={}& c_0 e^{-\kappa t} + c_1 e^{-\left(\kappa+\sqrt{\delta}\right)t} + c_2 e^{-\left(\kappa-\sqrt{\delta}\right)t} \\
& + c_3 e^{-\frac{1}{2}\left(\kappa+\sqrt{\delta}\right)t} + c_4 e^{-\frac{1}{2}\left(\kappa-\sqrt{\delta}\right)t} \\
&+ \frac{\Omega^2 \kappa_c^2}{\left(g^2+\kappa_c \kappa_m\right)^2}.
\end{aligned}\label{solutions3}
\end{equation}
Apparently, it contains the same five characteristic channels as that of the zero-photon initial state, shown in Eq.~\eqref{solutions2}, except that the explicit forms of the coefficients $c_i$, $i=0,1,2,3,4$, are different and are given in detail in Appendix \ref{AppB}. This indicates that the universality of the characteristic decay channels.

An experimentally feasible approach to observe this phenomenon can be achieved through two sequential steps: (1) Using a field with strength $\Omega_1$ to drive the cavity magnon system to a steady state from a vacuum state, i.e., $\langle m^{\dag} m \rangle=\langle c^{\dag} c \rangle=\langle c^{\dag} m \rangle=\langle m^{\dag} c \rangle=\langle m^{\dag} \rangle'=\langle m \rangle'=\langle c^{\dag} \rangle'=\langle c \rangle'=0$, of which the dynamics of magnon number $\left\langle m^{\dagger} m\right\rangle (t)$ can be obtained from Eq.~\eqref{solutions3} with $N_m=N_c=0$; (2) Reducing the driving field strength to $\Omega_2 < \Omega_1$ while maintaining other parameters constant.

Step 1 is illustratively shown in Fig.~\ref{drive_from_vacuum} region 1 ($t\le 400$ ns), where magnon numbers are increasing rapidly from zero and approaching to a steady state for different coupling strengths $g$.
\begin{figure}[htbp]
    \centering    \includegraphics[width=\columnwidth]{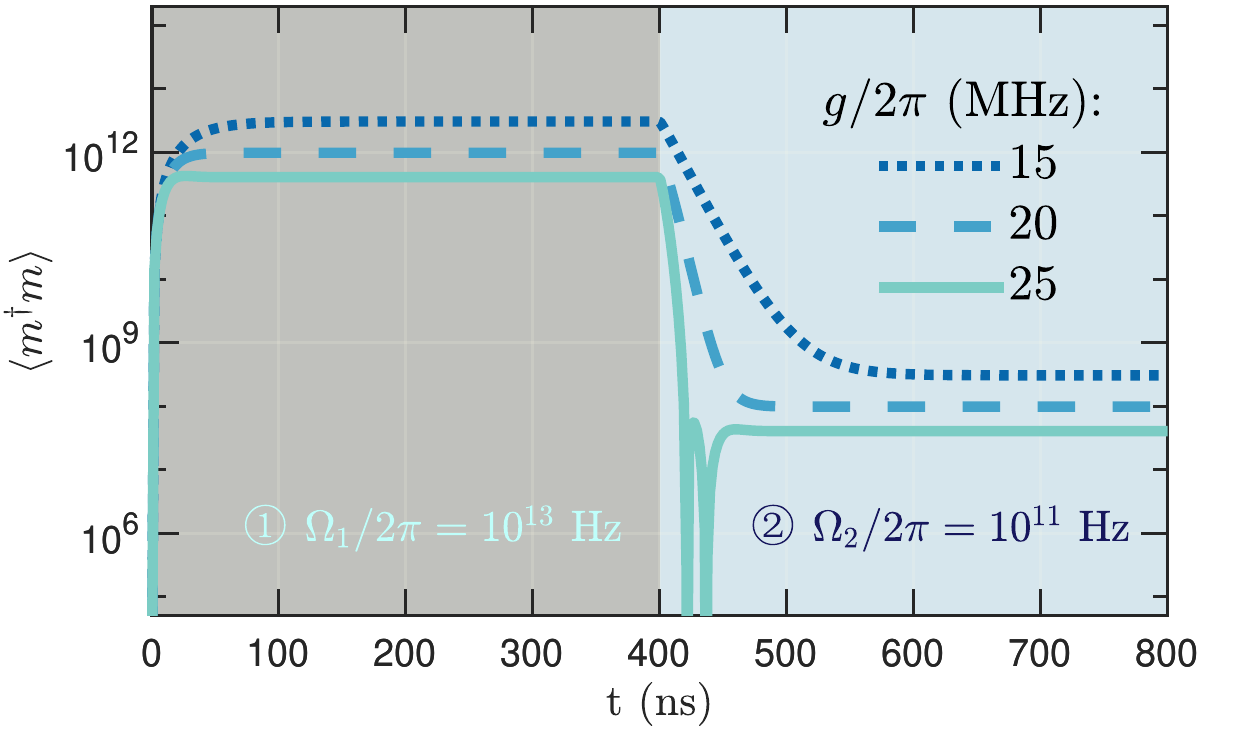}
    \caption{The magnons dynamics with three typical coupling rates, respectively. In region 1, both magnons and photons are driven from vacuum state at 0 ns with drive strength $\Omega_1/2\pi=10^{13}$ Hz. In region 2, drive strength is changed to $\Omega_2/2\pi=10^{11}$ Hz from 400 ns to 800 ns. The other parameters in both regions remain the same as in Fig.~\ref{decay_with_drive} (d).}\label{drive_from_vacuum}
\end{figure}

Step 2 is shown in Fig.~\ref{drive_from_vacuum} region 2, which represents the decay process from one steady state to another (lower) steady state. The qualitative features of this dynamical decay process for different coupling strengths are similar to that of the zero-photon initial state case which is shown Fig.~\ref{decay_with_drive} (d). This result directly indicates that the Purcell effect is a general effect for different initial states.
}

\section{Conclusion}
In summary, we have systematically analyzed cavity effects on the dynamics of magnon excitation in a general cavity-magnonic open system, considering both driven and non-driven cases. We present a comprehensive method for modeling multi-level cavity-coupled oscillator decay using coupled dynamical equations for magnon and cavity photon operators derived through the master equation approach. Analytical expressions for the Purcell effect on the magnon decay rate are obtained as functions of the magnon-cavity coupling strength $g$, detuning $\Delta$, and external drive amplitude $\Omega$ in the resonant case. Our theoretical predictions for the decay of the cavity photon number show excellent agreement with previous experimental results, validating the robustness of our model.

\revision{We identify four fundamental decay channels for the cavity-magnonic system in the absence of external driving. An additional four drive-controlled decay channels are found with driving fields for two complete different initial states. These channels represent the key decay characteristics of cavity-magnon open system. Furthermore, our analytical results explicitly delineate the pure and oscillatory decay regimes, naturally defining the weak coupling regime $0<g\le(\kappa_c-\kappa_m)/2$ in contrast to strong coupling $g>(\kappa_c-\kappa_m)/2$.}

\revision{Our framework applies to general two-oscillator coupling systems under negligible thermal noise and weak interaction (enabling use of RWA). These conditions align with most real-world implementations such as optomechanical systems, multi-mode cavities, cavity magnon-phonon hybrid systems, etc. This approach may also offer valuable insights for advancing quantum control, quantum device fabrication, and quantum resource preservation in magnonic systems and related quantum cavity platforms.}\\

\noindent{\bf Acknowledgment:} GZ and XQ acknowledge support from NSF Grant No. PHY-2316878. YW acknowledges support from NSFC No. 12074195.

\noindent{\bf Data availability:}  The numerical data that support the findings of this article are openly available \cite{data}. Due to size restriction, the two-dimensional surface plot data in Fig. \ref{decay_with_g} (a) and Fig. \ref{decay_with_drive} (c) is available upon request to the authors.



\appendix

{
\section{Equations of Motion without Drive}
\label{AppA}


The equations of motion \eqref{NumEq} for the non-drive case in the main text can be achieved through the master equation \eqref{masterequ}, i.e., 
\begin{widetext}
\begin{align}
\frac{d\left\langle c^{\dag} c\right\rangle}{d t} ={}& -i \omega_c \operatorname{Tr}\left[c^{\dag} c c^{\dag} c \rho-c^{\dag} c \rho c^{\dag} c\right]-i \omega_m \operatorname{Tr}\left[c^{\dag} c m^{\dag} m \rho-c^{\dag} c \rho m^{\dag} m \right] \notag\\
& -i g \operatorname{Tr}\left[ c^{\dag} c m^{\dag} c \rho-c^{\dag} c \rho m^{\dag} c \right] -i g \operatorname{Tr}\left[ c^{\dag} c c^{\dag} m \rho-c^{\dag} c \rho c^{\dag} m \right] \notag\\
& -\kappa_c \operatorname{Tr}\left[ c^{\dag} c c^{\dag} c \rho-2 c^{\dag} c c \rho c^{\dag}+c^{\dag} c \rho c^{\dag} c \right] \notag\\
& -\kappa_m \operatorname{Tr}\left[ c^{\dag} c m^{\dag} m \rho-2 c^{\dag} c m \rho m^{\dag}+c^{\dag} c \rho m^{\dag} m \right] \notag\\
={}& -i g \langle c^{\dag} m \rangle+i g \langle m^{\dag} c \rangle-2 \kappa_c \langle c^{\dag} c \rangle, \label{eqs1}\\
\frac{d\left\langle m^{\dag} m\right\rangle}{d t} ={}& -i \omega_c \operatorname{Tr}\left[m^{\dag} m c^{\dag} c \rho-m^{\dag} m \rho c^{\dag} c\right]-i \omega_m \operatorname{Tr}\left[m^{\dag} m m^{\dag} m \rho-m^{\dag} m \rho m^{\dag} m \right] \notag\\
& -i g \operatorname{Tr}\left[ m^{\dag} m m^{\dag} c \rho-m^{\dag} m \rho m^{\dag} c \right] -i g \operatorname{Tr}\left[ m^{\dag} m c^{\dag} m \rho-m^{\dag} m \rho c^{\dag} m \right] \notag\\
& -\kappa_c \operatorname{Tr}\left[ m^{\dag} m c^{\dag} c \rho-2 m^{\dag} m c \rho c^{\dag}+m^{\dag} m \rho c^{\dag} c \right] \notag\\
& -\kappa_m \operatorname{Tr}\left[ m^{\dag} m m^{\dag} m \rho-2 m^{\dag} m m \rho m^{\dag}+m^{\dag} m \rho m^{\dag} m \right] \notag\\
={}& i g \langle c^{\dag} m \rangle-i g \langle m^{\dag} c \rangle-2 \kappa_m \langle m^{\dag} m \rangle, \\
\frac{d\left\langle c^{\dag} m\right\rangle}{d t} ={}& -i \omega_c \operatorname{Tr}\left[c^{\dag} m c^{\dag} c \rho-c^{\dag} m \rho c^{\dag} c\right]-i \omega_m \operatorname{Tr}\left[c^{\dag} m m^{\dag} m \rho-c^{\dag} m \rho m^{\dag} m \right] \notag\\
& -i g \operatorname{Tr}\left[ c^{\dag} m m^{\dag} c \rho-c^{\dag} m \rho m^{\dag} c \right] -i g \operatorname{Tr}\left[ c^{\dag} m c^{\dag} m \rho-c^{\dag} m \rho c^{\dag} m \right] \notag\\
& -\kappa_c \operatorname{Tr}\left[ c^{\dag} m c^{\dag} c \rho-2 c^{\dag} m c \rho c^{\dag}+c^{\dag} m \rho c^{\dag} c \right] \notag\\
& -\kappa_m \operatorname{Tr}\left[ c^{\dag} m m^{\dag} m \rho-2 c^{\dag} m m \rho m^{\dag}+c^{\dag} m \rho m^{\dag} m \right] \notag\\
={}& i \omega_c \langle c^{\dag} m \rangle-i \omega_m \langle c^{\dag} m \rangle-i g \langle c^{\dag} c \rangle+i g \langle m^{\dag} m \rangle-\kappa_c \langle c^{\dag} m \rangle-\kappa_m \langle c^{\dag} m \rangle, \\
\frac{d\left\langle m^{\dag} c \right\rangle}{d t} ={}& -i \omega_c \operatorname{Tr}\left[m^{\dag} c c^{\dag} c \rho-m^{\dag} c \rho c^{\dag} c\right]-i \omega_m \operatorname{Tr}\left[m^{\dag} c m^{\dag} m \rho-m^{\dag} c \rho m^{\dag} m \right] \notag\\
& -i g \operatorname{Tr}\left[ m^{\dag} c m^{\dag} c \rho-m^{\dag} c \rho m^{\dag} c \right] -i g \operatorname{Tr}\left[ m^{\dag} c c^{\dag} m \rho-m^{\dag} c \rho c^{\dag} m \right] \notag\\
& -\kappa_c \operatorname{Tr}\left[ m^{\dag} c c^{\dag} c \rho-2 m^{\dag} c c \rho c^{\dag}+m^{\dag} c \rho c^{\dag} c \right] \notag\\
& -\kappa_m \operatorname{Tr}\left[ m^{\dag} c m^{\dag} m \rho-2 m^{\dag} c m \rho m^{\dag}+m^{\dag} c \rho m^{\dag} m \right] \notag\\
={}& -i \omega_c \langle m^{\dag} c \rangle+i \omega_m \langle m^{\dag} c \rangle+i g \langle c^{\dag} c \rangle-i g \langle m^{\dag} m \rangle-\kappa_c \langle m^{\dag} c \rangle-\kappa_m \langle m^{\dag} c \rangle, \label{eqs4}
\end{align}
\end{widetext}
where we have used the commutation relation $[a,a^{\dag}]=1$ and $[m,m^{\dag}]=1$. The compact form \eqref{NumEq} is just a straightforward combination of the above four equations \eqref{eqs1}-\eqref{eqs4}.

\section{Magnon Number Dynamical Equation with Drive}
\label{AppB}
The equations of motion \eqref{NumEqD} with drive in the main text can be solved by transforming the coefficient matrix $\mathcal{D}$ to the diagonal form $P\mathcal{D}P^{-1}$, where $P^{-1}$ is the matrix composed of the eigenvectors of $\mathcal{D}$. 

Then the equations for these eigen modes, denoted as $\la I_i(t)\ra$ ($i=1,2,...,8$), are achieved as $d\left\langle I_i(t) \right\rangle/ d t = \Gamma_i \left\langle I_i(t) \right\rangle + G_i$ with the solution $\left\langle I_i(t) \right\rangle = C_i e^{\Gamma_i t} - G_i/\Gamma_i$. Here $C_i$ is the constant determined by initial conditions, $\Gamma_i$ are the eigenvalues of $\mathcal{D}$, given in Eq.~\eqref{coefficient2} of the main text, and
\begin{widetext}
\begin{equation}
    ( G_1, G_2, G_3, G_4, G_5, G_6, G_7, G_8 )^T = P( 0,0,0,0,0,0,\Omega,\Omega )^T,
\end{equation}
and 
\begin{align}
&&( \left\langle I_1(t) \right\rangle, \left\langle I_2(t) \right\rangle, \left\langle I_3(t) \right\rangle, \left\langle I_4(t) \right\rangle, \left\langle I_5(t) \right\rangle, \left\langle I_6(t) \right\rangle, \left\langle I_7(t) \right\rangle, \left\langle I_8(t) \right\rangle )^T \notag \\
&&= P\big( \left\langle c^{\dag} c\right\rangle, \left\langle m^{\dag} m\right\rangle, \left\langle c^{\dag} m\right\rangle, \left\langle m^{\dag} c \right\rangle, \left\langle c^{\dag} \right\rangle ', \left\langle c \right\rangle ', \left\langle m^{\dag} \right\rangle ', \left\langle m \right\rangle ' \big)^T. \label{ModeTransformation}
\end{align}

When at resonance, i.e., $\omega_c=\omega_m=\omega_d$, there are five eigenvalues due to degeneracy
\begin{align}
\gamma_0&=\Gamma_1=\Gamma_2=-\kappa, \\
\gamma_1&=\Gamma_3=-\left(\kappa+\sqrt{\delta}\right), \\
\gamma_2&=\Gamma_4=-\left(\kappa-\sqrt{\delta}\right), \\
\gamma_3&=\Gamma_5=\Gamma_7=-\frac{1}{2} \left(\kappa + \sqrt{\delta}\right), \\
\gamma_4&=\Gamma_6=\Gamma_8=-\frac{1}{2} \left(\kappa - \sqrt{\delta}\right).
\end{align}
Therefore the final solution of average magnon number takes the form
\begin{equation}
\left\langle m^{\dag} m \right\rangle(t)=\sum^4_{i=0} b_i e^{\gamma_i t} + {\rm constant}. 
\end{equation}
For the initial coherent state discussed in section \ref{section3}, the coefficients $b_i$ can be determined explicitly as
\begin{align}
b_0 ={}& \frac{-2 g^2 N_m}{\delta}-\frac{2 g^2 \Omega\left(\Omega-\kappa\sqrt{N_m}\right)}{\delta\left(g^2+\kappa_c \kappa_m\right)}, \\
b_1 ={}& \frac{\left(d-\sqrt{\delta}\right)^2 N_m}{4\delta}-\frac{2 g^2 \Omega^2\left(2 \sqrt{\delta} g^2-\kappa_c\left(\sqrt{\delta} d-\delta\right) \right) + 4 g^2 \Omega \sqrt{N_m} \left(\sqrt{\delta} d-\delta\right) \left(g^2+\kappa_c \kappa_m\right)}{\sqrt{\delta} \delta\left(d+\sqrt{\delta}\right)\left(\kappa+\sqrt{\delta}\right)\left(g^2+\kappa_c \kappa_m\right)}, \\
b_2 ={}& \frac{\left(d+\sqrt{\delta}\right)^2 N_m}{4\delta}-\frac{2 g^2 \Omega^2\left(2 \sqrt{\delta} g^2-\kappa_c\left(\sqrt{\delta} d+\delta\right) \right) + 4 g^2 \Omega \sqrt{N_m} \left(\sqrt{\delta} d+\delta\right) \left(g^2+\kappa_c \kappa_m\right)}{\sqrt{\delta} \delta\left(d-\sqrt{\delta}\right)\left(\kappa-\sqrt{\delta}\right)\left(g^2+\kappa_c \kappa_m\right)}, \\
b_3 ={}& \frac{\Omega \kappa_c(d-\sqrt{\delta})\left(2 \Omega-\sqrt{N_m}\left(\kappa+\sqrt{\delta}\right)\right)}{\sqrt{\delta}\left(\kappa+\sqrt{\delta}\right)\left(g^2+\kappa_c \kappa_m\right)}, \\
b_4 ={}& \frac{\Omega \kappa_c(d+\sqrt{\delta})\left(-2 \Omega+\sqrt{N_m}\left(\kappa-\sqrt{\delta}\right)\right)}{\sqrt{\delta}\left(\kappa-\sqrt{\delta}\right)\left(g^2+\kappa_c \kappa_m\right)},
\end{align}
and the constant term, which is determined by \eqref{ModeTransformation}, represents the steady state excitation number when $t\rightarrow \infty$. Those constant terms of steady state are independent of the initial states and are formulated as
\begin{align}
\left\langle c^{\dag} c\right\rangle(\infty)={}& \frac{\Omega^2 g^2}{\left(g^2+\kappa_c \kappa_m\right)^2}, \\
\left\langle m^{\dag} m\right\rangle(\infty)={}& \frac{\Omega^2 \kappa_c^2}{\left(g^2+\kappa_c \kappa_m\right)^2}, \\
\left\langle c^{\dag} m\right\rangle^*(\infty)={}& \left\langle m^{\dag} c \right\rangle(\infty)= \frac{i g \kappa_c \Omega^2}{\left(g^2+\kappa_c \kappa_m\right)^2}, \\
\left\langle c^{\dag} \right\rangle '^*(\infty)={}& \left\langle c \right\rangle '(\infty)= \frac{i g \Omega}{g^2+\kappa_c \kappa_m}, \\
\left\langle m^{\dag} \right\rangle '(\infty)={}& \left\langle m \right\rangle '(\infty)= \frac{\kappa_c \Omega}{g^2+\kappa_c \kappa_m},
\end{align}
which hold the coherent state properties.

For an initial steady state used in section \ref{section4}, the solution of average magnon number takes the form
\begin{equation}
\left\langle m^{\dag} m \right\rangle(t)=\sum^4_{i=0} c_i e^{\gamma_i t} + \frac{\Omega^2 \kappa_c^2}{\left(g^2+\kappa_c \kappa_m\right)^2},
\end{equation}
with the coefficients
\begin{align}
c_0 ={}& \frac{-2 g^2 N_m-2 g^2 N_c-2 gd \sqrt{N_m N_c}}{\delta}-\frac{2 g^2\Omega^2 - 2g^2\Omega\kappa\sqrt{N_m} + 2g\Omega\sqrt{N_c}\left(2g^2-\kappa_c d\right)}{\delta\left(g^2+\kappa_c \kappa_m\right)}, \\
c_1 ={}& \frac{\left(\left(d-\sqrt{\delta}\right)\sqrt{N_m} + 2g\sqrt{N_c}\right)^2}{4\delta}-\frac{2 g^2 \Omega^2\left(2 \sqrt{\delta} g^2-\kappa_c\left(\sqrt{\delta} d-\delta\right) \right) + 4 g^2 \Omega \sqrt{N_m} \left(\sqrt{\delta} d-\delta\right) \left(g^2+\kappa_c \kappa_m\right)}{\sqrt{\delta} \delta\left(d+\sqrt{\delta}\right)\left(\kappa+\sqrt{\delta}\right)\left(g^2+\kappa_c \kappa_m\right)} \notag\\
& + \frac{g\Omega\sqrt{N_c}\left(2g^2-\kappa_c\left(d-\sqrt{\delta}\right)\right)}{\delta\left(g^2+\kappa_c \kappa_m\right)}, \\
c_2 ={}& \frac{\left(\left(d+\sqrt{\delta}\right)\sqrt{N_m} + 2g\sqrt{N_c}\right)^2}{4\delta}-\frac{2 g^2 \Omega^2\left(2 \sqrt{\delta} g^2-\kappa_c\left(\sqrt{\delta} d+\delta\right) \right) + 4 g^2 \Omega \sqrt{N_m} \left(\sqrt{\delta} d+\delta\right) \left(g^2+\kappa_c \kappa_m\right)}{\sqrt{\delta} \delta\left(d-\sqrt{\delta}\right)\left(\kappa-\sqrt{\delta}\right)\left(g^2+\kappa_c \kappa_m\right)} \notag\\
& + \frac{g\Omega\sqrt{N_c}\left(2g^2-\kappa_c\left(d+\sqrt{\delta}\right)\right)}{\delta\left(g^2+\kappa_c \kappa_m\right)}, \\
c_3 ={}& \frac{\Omega \kappa_c\left(d-\sqrt{\delta}\right)\left(2 \Omega-\sqrt{N_m}\left(\kappa+\sqrt{\delta}\right)\right)-2g\Omega\kappa_c \sqrt{N_c} \left(\kappa+\sqrt{\delta}\right)}{\sqrt{\delta}\left(\kappa+\sqrt{\delta}\right)\left(g^2+\kappa_c \kappa_m\right)}, \\
c_4 ={}& \frac{\Omega \kappa_c\left(d+\sqrt{\delta}\right)\left(-2 \Omega+\sqrt{N_m}\left(\kappa-\sqrt{\delta}\right)\right)+2g\Omega\kappa_c \sqrt{N_c} \left(\kappa-\sqrt{\delta}\right)}{\sqrt{\delta}\left(\kappa-\sqrt{\delta}\right)\left(g^2+\kappa_c \kappa_m\right)}.
\end{align}
\end{widetext}

Apparently, those $c_i$s reduce to the forms of $b_i$s when $N_c=0$, and to the forms of $a_i$s when $N_c=0$ and $\Omega=0$. Furthermore, it is worth to note that for other initial states of the magnon, e.g., Fock states and entangled states, corresponding analytical solutions can also be achieved.
}

\bibliography{References}

\end{document}